\documentclass[a4paper]{jpconf}

\usepackage{citesort}
\usepackage{graphicx}

\begin{document}

\title{Multi-dimensional potential energy surfaces and non-axial octupole correlations
       in actinide and transfermium nuclei from relativistic mean field models}
\author{Bing-Nan Lu$^1$,
        Jie Zhao$^1$,
	En-Guang Zhao$^{1,2}$,
    	Shan-Gui Zhou$^{1,2}$}
\address{%
$^1$State Key Laboratory of Theoretical Physics, Institute of Theoretical
Physics, Chinese Academy of Sciences, Beijing 100190, China \\
$^2$Center of Theoretical Nuclear Physics, National Laboratory of Heavy
Ion Accelerator, Lanzhou 730000, China
}
\ead{sgzhou@itp.ac.cn (Shan-Gui Zhou)}

\begin{abstract}
We have developed multi-dimensional constrained covariant density functional theories 
(MDC-CDFT) for finite nuclei in which the shape degrees of freedom $\beta_{\lambda\mu}$ 
with even $\mu$, e.g., $\beta_{20}$, $\beta_{22}$, $\beta_{30}$,
$\beta_{32}$, $\beta_{40}$, etc., can be described simultaneously.
The functional can be one of the following four forms:
the meson exchange or point-coupling nucleon interactions combined with
the non-linear or density-dependent couplings.
For the pp channel, either the BCS approach or the Bogoliubov transformation is implemented. 
The MDC-CDFTs with the BCS approach for the pairing (in the following labelled as 
MDC-RMF models with RMF standing for ``relativistic mean field'')
have been applied to investigate multi-dimensional potential
energy surfaces and the non-axial octupole $Y_{32}$-correlations in $N=150$ isotones.
In this contribution we present briefly the formalism of MDC-RMF models and
some results from these models.
The potential energy surfaces with and without triaxial
deformations are compared and it is found that the triaxiality plays
an important role upon the second fission barriers of actinide nuclei.
In the study of $Y_{32}$-correlations in $N=150$ isotones,
it is found that, for $^{248}$Cf and $^{250}$Fm, $\beta_{32} > 0.03$ and 
the energy is lowered by the $\beta_{32}$ distortion by more than 300 keV; 
while for $^{246}$Cm and $^{252}$No, the pocket with respect to $\beta_{32}$ is quite shallow.
\end{abstract}

\section{Introduction}

\begin{figure}
\begin{center}
\resizebox{0.9\columnwidth}{!}{%
 \includegraphics{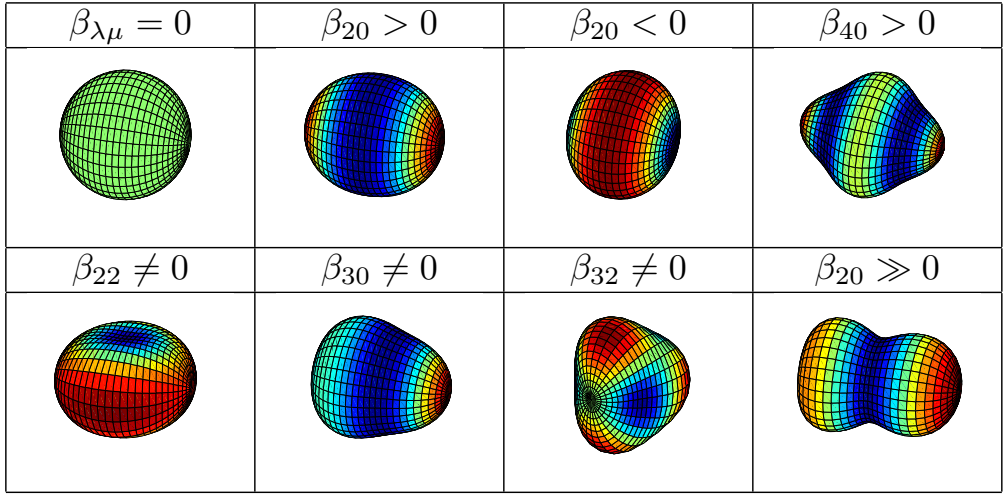} }
\end{center}
\caption{\label{Pic:deformations}(Color online)
A schematic show of some typical nuclear shapes. 
From left to right, the 1st row: (a) Sphere, (b) Prolate spheroid,
(c) Oblate spheroid, (d) Hexadecapole shape, 
and the second row: (e) Triaxial ellipsoid,
(f) Reflection symmetric octupole shape, (g) Tetrahedron,
(h) Reflection asymmetric octupole shape with very large quadrupole deformation
    and large hexadecapole deformation.
Taken from Ref.~\cite{Lu2012_PhD}.
}
\end{figure}

The shape of a nucleus can be described by the parametrization of the
nuclear surface or the nucleon density distribution~\cite{Bohr1998_Nucl_Structure_1,Ring1980}.
One of the mostly used parametrizations is the multipole expansion with 
coefficients $\beta_{\lambda\mu}$'s;
see Fig.~\ref{Pic:deformations}~\cite{Lu2012_PhD}
for a schematic show of some typical nuclear shapes.
Many shape degrees of freedom are important  
not only for the ground states or small amplitude collective motions, but also
for large amplitude collective motions such as fission.
The readers are referred to Ref.~\cite{Lu2013_in-prep} for an overview.
Here we simply mention that both the non-axial and the reflection symmetries
should be broken in the study of nuclear ground state properties and
potential energy surfaces.

We have developed multi-dimensional constrained covariant density functional theories (MDC-CDFT)
by breaking the reflection and the axial symmetries 
simultaneously~\cite{Lu2012_PRC85-011301R,Lu2012_PhD,%
Zhao2012_PRC86-057304,Lu2012_EPJWoC38-05003,Lu2013_in-prep}.
In these theories, 
the nuclear shape is assumed to be invariant under 
the reversion of $x$ and $y$ axes, i.e., 
the intrinsic symmetry group is $V_{4}$ and 
all shape degrees of freedom $\beta_{\lambda\mu}$ with even $\mu$, 
e.g., $\beta_{20}$, $\beta_{22}$, $\beta_{30}$, $\beta_{32}$, $\beta_{40}$, $\cdots$, 
are included self-consistently.
The covariant density functional can be one of the following four forms: 
the meson exchange or point-coupling nucleon interactions combined with 
the non-linear or density-dependent couplings.
For the pp channel, either the BCS approach or the Bogoliubov transformation is implemented.
The MDC-CDFT with the BCS approach for the pairing is
named as MDC-RMF and that with the Bogoliubov transformation as MDC-RHB~\cite{Lu2013_in-prep}.
In this contribution, we will present the formalism for MDC-RMF models 
and some results of actinide nuclei.

In Section~\ref{sec:formalism}, we give the formalism of MDC-RMF models briefly.
The results for actinide nuclei are shown and discussed in Section~\ref{sec:results}.
A summary is given in Section~\ref{sec:summary}.

\section{Formalism of MDC-RMF models}
\label{sec:formalism}

In this section, we briefly give the formalism of MDC-RMF models.
More details can be found in Ref.~\cite{Lu2013_in-prep}.
The starting point of a RMF model with the non-linear point coupling 
interactions is the following Lagrangian~\cite{Serot1986_ANP16-1,Reinhard1989_RPP52-439,%
Ring1996_PPNP37-193,Vretenar2005_PR409-101,Meng2006_PPNP57-470,Paar2007_RPP70-691,%
Niksic2011_PPNP66-519,%
Prassa2012_PRC86-024317,Meng2013_FPC8-55},
\begin{equation}
 \mathcal{L} = \bar{\psi} \left( i\gamma_{\mu}\partial^{\mu}-M_{B} \right) \psi
             - \mathcal{L}_{{\rm lin}} -\mathcal{L}_{{\rm nl}}
             - \mathcal{L}_{{\rm der}} -\mathcal{L}_{{\rm cou}},
\end{equation}
where
\begin{eqnarray}
 \mathcal{L}_{{\rm lin}} & = & \frac{1}{2} \alpha_{S}\rho_{S}^{2}
                              +\frac{1}{2}\alpha_{V} j_{V}^{2}
                              +\frac{1}{2}\alpha_{TS}\vec{\rho}_{TS}^{2}
                              +\frac{1}{2}\alpha_{TV}\vec{j}_{TV}^{2}
 ,
 \nonumber \\
 \mathcal{L}_{{\rm nl }} & = & \frac{1}{3}\beta_{S}  \rho_{S}^{3}
                              +\frac{1}{4}\gamma_{S} \rho_{S}^{4}
                              +\frac{1}{4}\gamma_{V} \left[j_{V}^{2}\right]^{2}
 ,
 \nonumber \\
 \mathcal{L}_{{\rm der}} & = & \frac{1}{2}\delta_{S} \left[\partial_{\nu}\rho_{S}\right]^{2}
                              +\frac{1}{2}\delta_{V} \left[\partial_{\nu}j_{V}^{\mu}\right]^{2}
                              +\frac{1}{2}\delta_{TS}\left[\partial_{\nu}\vec{\rho}_{TS}\right]^{2}
 \nonumber \\
                         &   &+\frac{1}{2}\delta_{TV}\left[\partial_{\nu}\vec{j}_{TV}\right]^{2}
 ,
 \nonumber \\
 \mathcal{L}_{{\rm cou}} & = & \frac{1}{4} F^{\mu\nu}F_{\mu\nu}
                              +e \frac{1-\tau_{3}}{2}A_{\mu}j_{V}^{\mu}
 ,
\end{eqnarray}
are the linear, non-linear, and derivative couplings and the Coulomb part, respectively.
$M_{{B}}$ is the nucleon mass, $\alpha_{S}$, $\alpha_{V}$, $\alpha_{TS}$,
$\alpha_{TV}$, $\beta_{S}$, $\gamma_{S}$, $\gamma_{V}$, $\delta_{S}$, $\delta_{V}$,
$\delta_{TS}$, and $\delta_{TV}$ are coupling constants for different channels and
$e$ is the electric charge.
$\rho_{S}$, $\vec{\rho}_{TS}$, $j_{V}$, and $\vec{j}_{TV}$ are the iso-scalar density,
iso-vector density, iso-scalar current, and iso-vector current, respectively.

Starting from the above Lagrangian and under several approximations, one can derive
the Dirac equation for the nucleons,
\begin{equation}
 \hat{h}\psi_{i} = 
 \left\{{\mathbf \alpha}\cdot\mathbf{p}+\beta[M_B+S(\mathbf{r})]+V(\mathbf{r})\right\}\psi_{i} = 
 \epsilon_{i}\psi_{i}
 ,
\end{equation}
where the potentials $V({\bf r})$ and $S({\bf r})$ are calculated from the densities.

An axially deformed harmonic oscillator (ADHO) basis is adopted for solving the
Dirac equation~\cite{Lu2012_PRC85-011301R,Lu2012_PhD,%
Zhao2012_PRC86-057304,Lu2012_EPJWoC38-05003,Lu2013_in-prep,Lu2011_PRC84-014328,%
Gambhir1990_APNY198-132,Ring1997_CPC105-77}.
Note that a RMF model with reflection asymmetry has been developed 
in a two-center HO basis~\cite{Geng2007_CPL24-1865}.
The ADHO basis are defined as the eigen solutions of the Schr\"odinger
equation with an ADHO potential,
\begin{eqnarray}
 \left[ -\frac{\hbar^{2}}{2M_B} \nabla^{2} + 
         \frac{1}{2} M_B ( \omega_{\rho}^{2}\rho^{2} + \omega_{z}^{2}z^{2} ) 
 \right] \Phi_{\alpha}(\bf{r}\sigma)
 & = &
 E_{\alpha} \Phi_{\alpha}(\bf{r}\sigma)
 ,
 \label{eq:BasSchrodinger-1}
\end{eqnarray}
where $\omega_{z}$ and $\omega_{\rho}$ are the oscillator frequencies along
and perpendicular to the $z$ axis, respectively.
These basis are also eigen functions of the $z$ component of the
angular momentum $j_{z}$ with eigen values $K=m_{l}+m_{s}$.
For any basis state $\Phi_{\alpha}(\bf{r}\sigma)$, the time reversal state
is defined as $\Phi_{\bar{\alpha}}(\bf{r}\sigma)=\mathcal{T}\Phi_{\alpha}(\bf{r}\sigma)$,
where $\mathcal{T}=i\sigma_{y}K$ is the time reversal operator and
$K$ is the complex conjugation.
Apparently we have $K_{\bar{\alpha}}=-K_{\alpha}$
and $\pi_{\bar{\alpha}}=\pi_{\alpha}$ with $\pi=\pm 1$ being the parity.
These basis form a complete set for expanding any two-component spinors.
For a Dirac spinor with four components,
\begin{equation}
 \psi_{i}(\bf{r}\sigma) =
 \left( \begin{array}{c}
        \sum_{\alpha}f_{i}^{\alpha} \Phi_{\alpha}(\bf{r}\sigma) \\
        \sum_{\alpha}g_{i}^{\alpha} \Phi_{\alpha}(\bf{r}\sigma)
        \end{array}
 \right),
\end{equation}
where the sum runs over all the possible combination of the quantum
numbers $\alpha=\{n_{z},n_{r},m_{l},m_{s}\}$ and $f_{i}^{\alpha}$ and
$g_{i}^{\alpha}$ are the expansion coefficients.
In practical calculations, one should truncate the basis in an effective way.

In our model the nucleus is assumed to be symmetric under the $V_4$ group, that is,
for all the potentials and densities we can do the Fourier series expansion,
\begin{equation}
 f(\rho,\varphi,z) =  f_{0}(\rho,z) \frac{1}{\sqrt{2\pi}}
+ \sum_{n=1}^{\infty} f_{n}(\rho,z) \frac{1}{\sqrt{\pi}}\cos(2n\varphi),
\end{equation}
where $f_0 (\rho, z)$ and $f_n (\rho, z)$ are real functions of $\rho$ and $z$.
The formalism for calculating the Fourier components of the potentials and
densities can be found in Ref.~\cite{Lu2013_in-prep}.

Either the BCS approach or the Bogoliubov transformation has been implemented
in our model to take into account the pairing effects.
For the pairing force, we can use a delta force or a separable finite-range
pairing force~\cite{Tian2009_PLB676-44,Tian2009_PRC80-024313,Niksic2010_PRC81-054318}.

To obtain a potential energy surface, i.e., the energy of a nucleus as a function of
deformation parameters, we make multi-dimensional constraint calculations
which are equivalent to adding external potentials during the iteration.
A modified linear constraint method was included in our MDC-RMF 
calculations~\cite{Lu2012_PRC85-011301R,Lu2012_PhD,%
Zhao2012_PRC86-057304,Lu2012_EPJWoC38-05003,Lu2013_in-prep}.

The total energy of a nucleus is obtained by substituting the densities
into the expectation value of the Hamiltonian.
The center of mass correction $E_{{\rm c.m.}}$ can be calculated
either phenomenologically or microscopically, depending on the effective interactions.
The intrinsic multipole moments are calculated from the vector densities by
\begin{equation}
 Q_{\lambda\mu} = \int d^{3}\mathbf{r} \rho_{V} (\mathbf{r})r^{\lambda}Y_{\lambda\mu}(\Omega),
\end{equation}
where $Y_{\lambda\mu}(\Omega)$ is the spherical harmonics.

\section{Results and discussions}
\label{sec:results}

\begin{figure}
\begin{center}
\resizebox{0.8\columnwidth}{!}{%
 \includegraphics{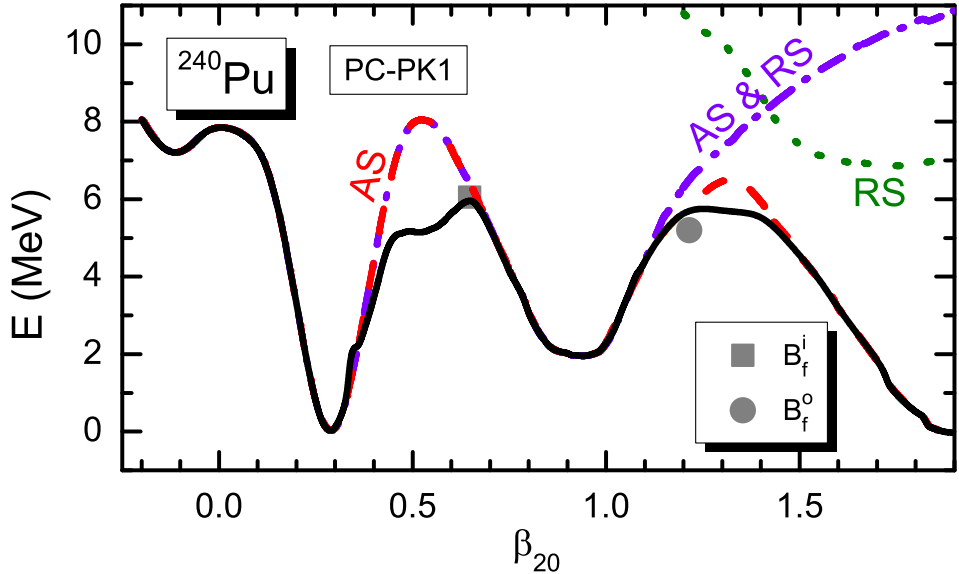} }
\end{center}
\caption{\label{fig:Pu240PES1d} (Color online)
Potential energy curve of $^{240}$Pu with various self-consistent symmetries imposed.
The solid black curve represents the calculated fission path with $V_4$ symmetry imposed,
the red dashed curve that with the axial symmetry (AS) imposed,
the green dotted curve that with the reflection symmetry (RS) imposed,
the violet dot-dashed line that with both symmetries (AS \& RS) imposed.
The empirical inner (outer) barrier height is taken from 
Ref.~\cite{Capote2009_NDS110-3107} and denoted by the grey square (circle).
The energy is normalized with respect to the binding energy of the ground state.
The parameter set used is PC-PK1.
Taken from Ref.~\cite{Lu2012_PRC85-011301R}.
}
\end{figure}

In Ref.~\cite{Lu2012_PRC85-011301R}, one- (1-d), two- (2-d), and three-dimensional (3-d)
constraint calculations were made for the actinide nucleus $^{240}$Pu using the MDC-RMF models 
with the parameter set PC-PK1~\cite{Zhao2010_PRC82-054319,Zhao2012_PRC86-064324}.
In Fig.~\ref{fig:Pu240PES1d} we show the 1-d potential energy curve
from an oblate shape with $\beta_{20}$ about $-0.2$ to the fission
configuration with $\beta_{20}$ beyond 2.0
which are obtained from calculations with different self-consistent symmetries
imposed: The axial (AS) or triaxial (TS) symmetries combined with
reflection symmetric (RS) or asymmetric cases.
The importance of the triaxial deformation on the inner barrier~\cite{Abusara2010_PRC82-044303,%
Afanasjev2013_arXiv1303.1206} 
and that of the octupole deformation on the outer barrier 
are clearly seen:
The triaxial deformation reduces the inner barrier height by more than 2 MeV
and results in a better agreement with the empirical value~\cite{Capote2009_NDS110-3107};
the RA shape is favored beyond the fission isomer and lowers very much
the outer fission barrier.
Besides these features, it was found for the first time that the outer
barrier is also considerably lowered by about 1 MeV when the triaxial
deformation is allowed.
In addition, a better reproduction of the empirical barrier height can be seen for
the outer barrier.
It has been stressed that this feature can only be found when the axial and
reflection symmetries are simultaneously broken~\cite{Lu2012_PRC85-011301R}.

\begin{figure}
\begin{center}
\resizebox{0.8\columnwidth}{!}{%
 \includegraphics{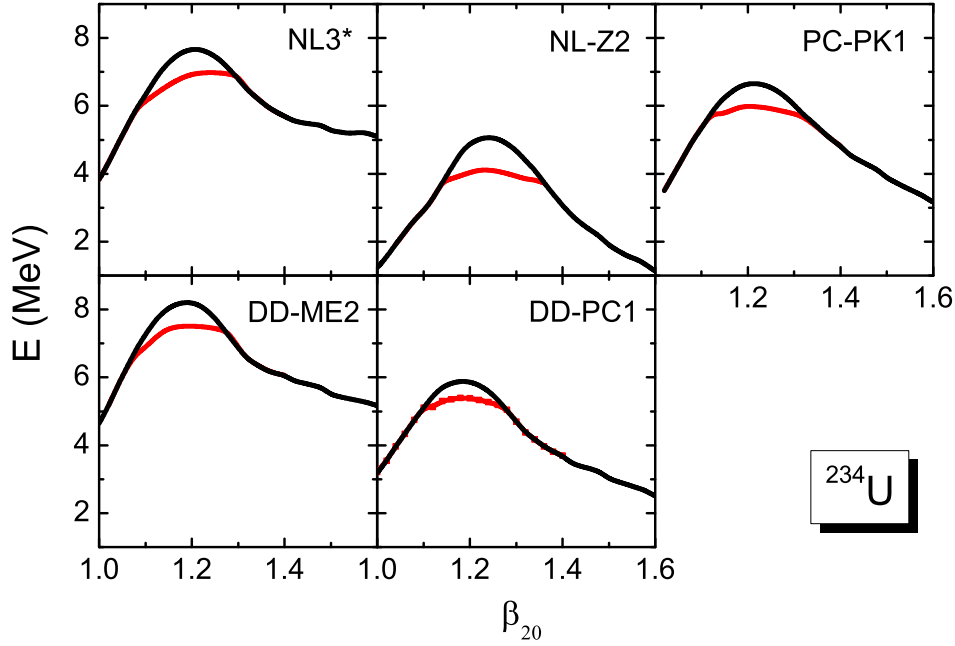} }
\end{center}
\caption{\label{fig:parameters} (Color online)
Potential energy curves of $^{234}$U in the second barrier region calculated from
MDC-RMF models with the NL3*~\cite{Lalazissis1997_PRC55-540,Lalazissis2009_PLB671-36},
NL-Z2~\cite{Bender1999_PRC60-034304}, PC-PK1~\cite{Zhao2010_PRC82-054319,Zhao2012_PRC86-064324},
DD-ME2~\cite{Lalazissis2005_PRC71-024312}, and DD-PC1~\cite{Niksic2008_PRC78-034318}
parameter set, respectively.
The reflection asymmetric shapes are allowed.
The axial and non-axial symmetric results are denoted by black and red curves, respectively.
Taken from Ref.~\cite{Lu2012_PhD}.
}
\end{figure}

In Ref.~\cite{Lu2012_PRC85-011301R}, it was also examined the parameter dependence
of the influence of triaxiality on the second fission barrier and
the lowering effect of the triaxiality on the second fission barrier
was also observed when parameter sets other than PC-PK1 are used.
In Fig.~\ref{fig:parameters} we show the comparison of the potential energy curves of
$^{234}$U in the second barrier region calculated from the MDC-RMF models with different
parameter sets and Lagrangian forms, 
including meson-exchange ones 
NL3{*}~\cite{Lalazissis1997_PRC55-540,Lalazissis2009_PLB671-36}, 
NL-Z2~\cite{Bender1999_PRC60-034304}, 
DD-ME2~\cite{Lalazissis2005_PRC71-024312}, and 
point-coupling ones PC-PK1~\cite{Zhao2010_PRC82-054319,Zhao2012_PRC86-064324} and 
DD-PC1~\cite{Niksic2008_PRC78-034318}.
The results with and without triaxiality are both presented.
From the figure we can see that,
although the absolute values of the barriers differ a lot among these results,
for all five parameter sets the second barriers are lowered by the triaxiality.
The largest effect is around 1 MeV for NL-Z2 and the smallest one is around 300 keV for DD-PC1.
Thus we conclude that the lowering effects of the triaxiality on the second barriers
are parameter independent in the RMF models.

\begin{figure}
\begin{center}
\resizebox{0.95\columnwidth}{!}{%
 \includegraphics{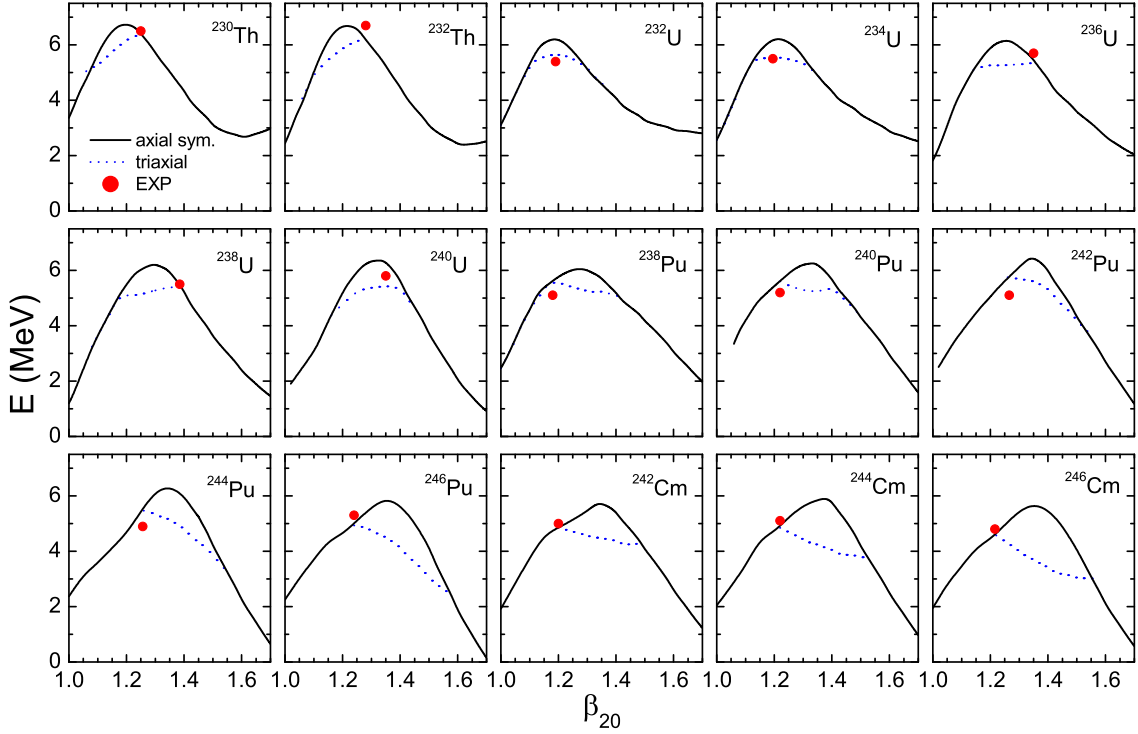} }
\end{center}
\caption{\label{fig:outerbarriers} (Color online)
Potential energy curves of even-even actinides nuclei in the second barrier regions calculated from 
the MDC-RMF model with the PC-PK1 parameter set~\cite{Zhao2010_PRC82-054319,Zhao2012_PRC86-064324}.
The reflection asymmetric shapes are allowed.
The axial and non-axial symmetric results are denoted by solid and dotted curves, respectively.
The binding energies are normalized with respect to the ground states.
The empirical values are taken from Ref.~\cite{Capote2009_NDS110-3107} and denoted by full circles.
Taken from Ref.~\cite{Lu2012_PhD}.
}
\end{figure}

The self-consistent three-dimensional constraint calculations are very time-consuming.
From the benchmark calculations for $^{240}$Pu, we learned many experiences about 
the important roles played by various shape degrees of freedom in different regions of 
the deformation space.
For example,
around the first fission barrier an actinide nucleus assumes triaxial and 
reflection symmetric shapes but around the second fission barrier both triaxial and 
octupole deformations are important~\cite{Lu2013_in-prep}.
These experiences are used in a systematic study of even-even actinide nuclei
and the results were presented in Ref.~\cite{Lu2013_in-prep}.
Here in Fig.~\ref{fig:outerbarriers} we show the 1-d potential energy curves for even-even
actinide nuclei from the fission isomer with $\beta_{20}=1.0$ to the fission configuration
with $\beta_{20}=1.7$ calculated from the MDC-RMF model 
with the parameter set PC-PK1~\cite{Zhao2010_PRC82-054319,Zhao2012_PRC86-064324}.
For comparison we present the results with and without the non-axial deformations.
The reflection asymmetric shapes are allowed in both calculations.
It was known that most of the outer barriers of the actinide nuclei are reflection asymmetric,
which is consistent with the observed low-energy asymmetric fission fragments in this region.
However, to estimate the fission barrier heights with an even higher accuracy,
it is desirable to include also the effects of the non-axial deformations.
From these potential energy curves it is clear that the non-axial deformations lower
the outer barriers by around 0.5 to 1 MeV compared with the axial symmetric results.
The empirical values of the fission barriers are also presented in the figure as full circles.
Apparently, by including the triaxiality, the agreement between the theory and
the experiment is improved.

\begin{figure}
\begin{center}
\resizebox{0.6\columnwidth}{!}{%
 \includegraphics{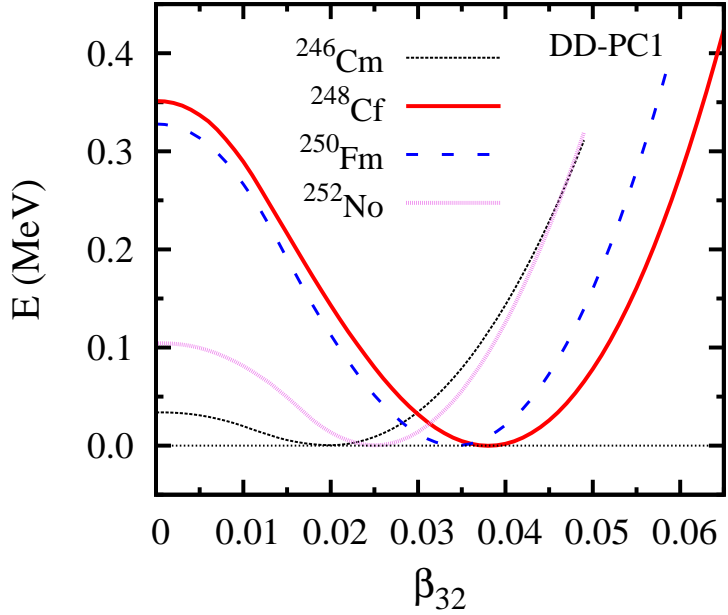} }
\end{center}
\caption{\label{fig:b32} (Color online)
The binding energy $E$ (relative to the ground state) for $N=150$ isotones $^{246}$Cm
(dashed line), $^{248}$Cf (solid line), $^{250}$Fm (long-dashed line),
and $^{252}$No (dotted line) as a function of the non-axial octupole
deformation parameter $\beta_{32}$.
Taken from Ref.~\cite{Zhao2012_PRC86-057304}.
}
\end{figure}

In Ref.~\cite{Zhao2012_PRC86-057304} the non-axial reflection-asymmetric $\beta_{32}$
shape in some transfermium nuclei with $N=150$, namely $^{246}$Cm, $^{248}$Cf,
$^{250}$Fm, and $^{252}$No were investigated using the MDC-RMF model.
In Fig.~\ref{fig:b32}, we show potential energy curves for these $N=150$ isotones.
For $^{246}$Cm, the ground state deformation $\beta_{32}=0.020$.
The potential energy curve is rather flat around the minimum.
We denote the energy difference between the ground state and the point with
$\beta_{32}=0$ by $E_\mathrm{depth}$ which
measures the energy gain with respect to the $\beta_{32}$ distortion.
For $^{246}$Cm, $E_\mathrm{depth}$ is only $34$ keV.
For $^{248}$Cf, $^{250}$Fm, and $^{252}$No, the minima locate at
$\beta_{32} = 0.037$, $0.034$, and $0.025$, respectively.
The corresponding energy gain $E_\mathrm{depth} = 0.351$, $0.328$, and $0.104$ MeV.
As is discussed in Ref.~\cite{Zhao2012_PRC86-057304},
the occurrence of the non-axial octupole $\beta_{32}$ correlations is mainly from a pair of
neutron orbitals $[734]9/2$ ($\nu j_{15/2}$) and $[622]5/2$
($\nu g_{9/2}$) which are close to the neutron Fermi surface and a pair of
proton orbitals $[521]3/2$ ($\pi f_{7/2}$) and $[633]7/2$ ($\pi i_{13/2}$)
which are close to the proton Fermi surface.

It may be hard to conclude that these nuclei have static non-axial
octupole deformations from these results because the potential energy curve
is flat around the minimum and $E_\mathrm{depth}$ is small~\cite{Zhao2012_PRC86-057304}.
However, the present calculations at least indicate a strong $Y_{32}$-correlation in these nuclei.
Both the non-axial octupole parameter $\beta_{32}$ and the energy gain $E_\mathrm{depth}$
reach maximal values at $^{248}$Cf in the four nuclei along the $N=150$ isotonic chain.
This is consistent with the analysis given in 
Refs.~\cite{Chen2008_PRC77-061305R,Jolos2011_JPG38-115103} and the experimental observation 
that in $^{248}$Cf, the $2^-$ state is the lowest among these 
nuclei~\cite{Robinson2008_PRC78-034308}.

\section{\label{sec:summary}Summary}

In this contribution we present the formalism and some applications of the multi-dimensional
constrained relativistic mean field (MDC-RMF) models in which all shape degrees of freedom
$\beta_{\lambda\mu}$ deformations with even $\mu$ are allowed.
The potential energy surfaces (curves) of $^{240}$Pu and actinide nuclei with various
symmetries are investigated.
It is found that besides the octupole deformation, the triaxiality also
plays an important role upon the second fission barriers.
For most of even-even actinide nuclei, the triaxiality lowers the outer barrier
by 0.5 $\sim$ 1 MeV, accounting for about 10 $\sim$ 20\% of the barrier height.
The non-axial reflection-asymmetric $\beta_{32}$ shape in some transfermium
nuclei with $N=150$, namely $^{246}$Cm, $^{248}$Cf, $^{250}$Fm, and $^{252}$No
are studied
and rather strong non-axial octupole $Y_{32}$ effects have been found in $^{248}$Cf and
$^{250}$Fm which are both well deformed with large axial-quadrupole deformations,
$\beta_{20} \approx 0.3$.
We note that it is crucial to include the reflection asymmetric and non-axial 
shapes simultaneously for the study of potential energy surfaces and fission barriers 
of actinide nuclei and of nuclei in unknown mass regions such as superheavy nuclei.

\section*{Acknowledgement}

This work has been supported by Major State Basic Research Development 
Program of China (Grant No. 2013CB834400), 
National Natural Science Foundation of China (Grant Nos. 11121403, 11175252, 
11120101005, 11211120152, and 11275248), 
Knowledge Innovation Project of Chinese Academy of Sciences (Grant Nos. KJCX2-
EW-N01 and KJCX2-YW-N32). 
The results described in this paper are obtained on the ScGrid of Supercomputing
Center, Computer Network Information Center of Chinese Academy of Sciences.

\section*{References}


\providecommand{\newblock}{}

\end{document}